\RequirePackage{ifpdf}
\documentclass{PoS}

\usepackage[authoryear,square]{natbib}
\bibpunct{(}{)}{;}{a}{}{,}

\title{Multi-wavelength, Multi-Messenger Pulsar Science in the SKA Era}

\ShortTitle{Multi-wavelength, Multi-Messenger Pulsar Science in the SKA Era}

\author{John Antoniadis,$^{1,2}$ Lucas Guillemot,$^3$ \speaker{Andrea Possenti},$^{, 4}$ Slavko Bogdanov,$^5$  Joseph~D.~Gelfand,$^6$ Michael Kramer,$^{1,7}$ Roberto Mignani,$^{8,9}$ Benjamin Stappers$^{7}$ \& Pablo Torne$^{1}$\\
		$^1$Max-Planck-Institut f\"{u}r Radioastronomie, Bonn, Germany; 
		$^2$Dunlap Institute for Astronomy and Astrophysics University of Toronto; 	
		$^3$Laboratoire de Physique et Chimie de l'Environnement, Orl\'{e}ans, France Station de radioastronomie de Nan\c{c}ay, Obs. de Paris, CNRS/INSU, Nan\c{c}ay, France; 
		$^4$ INAF-Osservatorio Astronomico di Cagliari, Selargius (CA), Italy; 
		$^5$ Columbia Astrophysics Laboratory, Columbia University, New York, USA; 
		$^6$ New York University, Abu Dhabi; 
		$^7$ Jodrell Bank Center for Astrophysics, University of Manchester, UK; 
		$^8$ INAF - Istituto di Astrofisica Spaziale e Fisica Cosmica Milano, Milano, Italy; 
		$^9$ Kepler Institute of Astronomy, University of Zielona G\'{o}ra, Zielona G\'{o}ra, Poland \\
		email: \email{antoniadis@dunlap.utoronto.ca}
		}

\abstract{
The Square Kilometre Array (SKA) is an integral part of the next-generation observatories that will survey the Universe across the electromagnetic spectrum, and 
beyond, revolutionizing our view of fundamental physics, astrophysics and cosmology. Owing to their extreme nature and clock-like properties, pulsars discovered and 
monitored by SKA will enable a broad range of scientific endeavour and play a key role in this quest. This chapter summarizes the pulsar-related science goals that will 
be reached with coordinated efforts among SKA and other next-generation astronomical facilities.
}

\FullConference{
Advancing Astrophysics with the Square Kilometre Array\\
June 8-13, 2014\\
Giardini Naxos, Italy}

\begin{document}

\section{Introduction}% (John)}

One of the key science goals for the SKA is to provide a nearly complete census of radio pulsars in the Milky Way and its Globular Clusters 
\citep{keane2014,hessels2014}. The large number of pulsar discoveries and the unprecedented timing precision of the instrument will enable a broad spectrum of science, 
ranging from characterization of stochastic gravitational wave (GW) signals \citep{janssen2014} to probing all possible outcomes of massive-star evolution \citep{tauris2014}.  

As  the SKA will be detecting its first light, a fleet of sensitive telescopes will be gathering photons at all wavelengths and second generation GW detectors will be 
making sensitive observations. This multi-wavelength, multi-messenger frontier in astronomy will allow to tackle many remaining open problems in neutron 
star (NS) physics, and study with unprecedented detail the Galactic structure and content, the nature of the strong interaction, strong-field gravity and the large-scale 
structure of the Universe. 

\begin{figure}[htp]
\begin{center}
\includegraphics[scale=0.36]{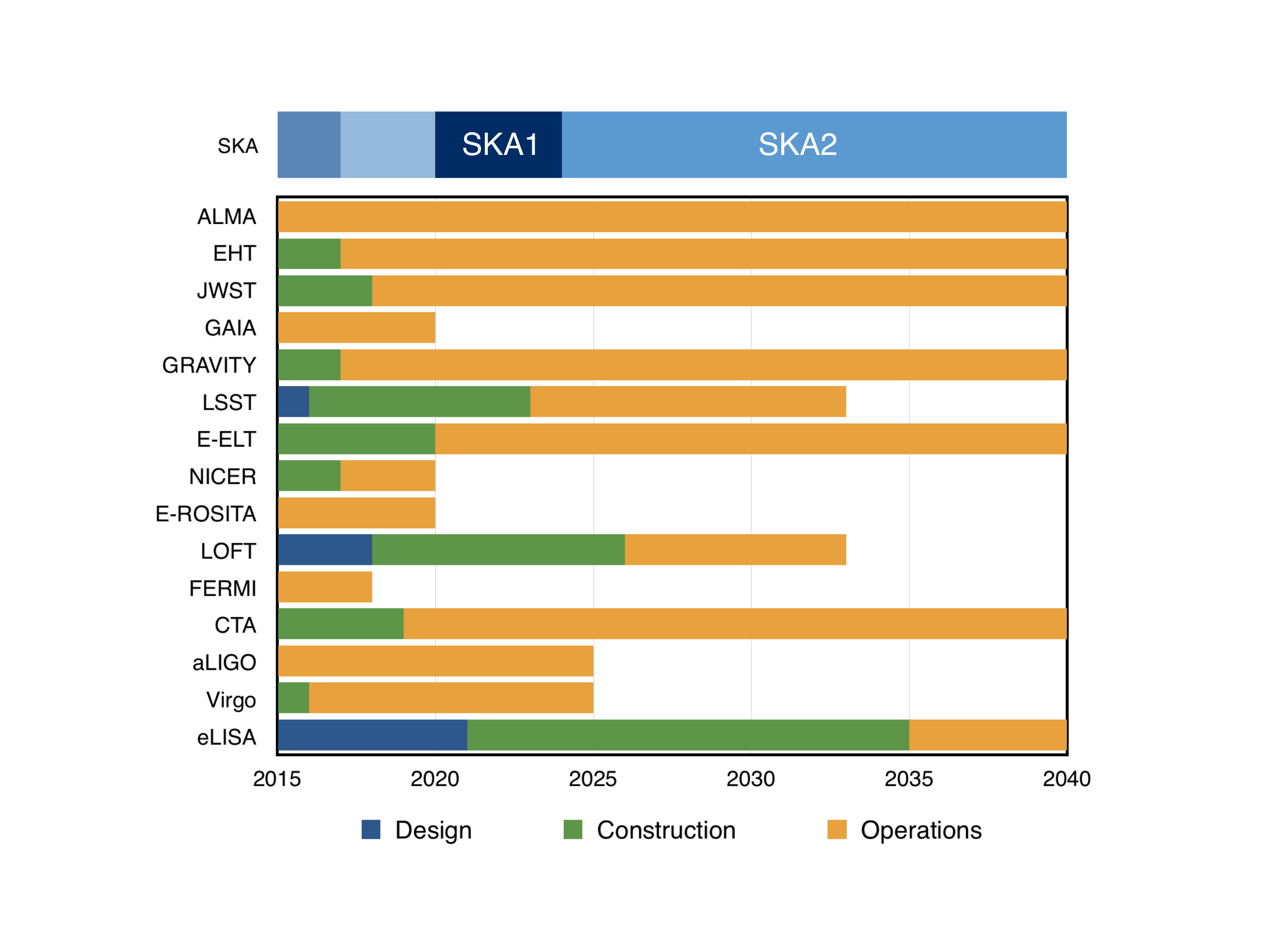} 
\label{fig1}
\caption{Approximate timeline for observatories of interest for synergistic neutron star science with the SKA.}
\end{center}
\end{figure}

Figure\,\ref{fig1} sketches an approximate timeline (design--construction--operations) for observatories that will be of particular importance for pulsar science. At radio and 
sub-mm wavelengths, ALMA\footnote{Atacama Large Millimeter/submillimeter Array, http://www.almaobservatory.org/} and EHT\footnote{Event-Horizon Telescope, 
http://www.eventhorizontelescope.org/}/BlackHoleCam\footnote{http://www.space.com/24002-black-hole-image-event-horizon.html}, will study the pulsar emission 
mechanism and magnetic structure and, jointly with pulsar-timing observations,  probe the nature of strong-field gravity around the super-massive black hole (SMBH) 
at the center of the Milky Way.  

Ground-based and space-borne  instruments such as GAIA\footnote{Global Astrometric Interferometer for Astrophysics, http://sci.esa.int/gaia/}, LSST\footnote{Large 
Synoptic Survey Telescope, http://www.lsst.org/lsst/}, JWST \footnote{James Webb Space Telescope, http://www.jwst.nasa.gov/} and E-ELT \footnote{The European 
Extremely Large Telescope, https://www.eso.org/public/teles-instr/e-elt/} will provide a thorough census of the Galaxy's stellar content, including the companions of 
pulsars discovered and monitored with the SKA. Similarly, studies of NS X-ray binaries with next-generation X-ray  telescopes such as NICER\footnote{Neutron star Interior Composition Explorer; http://heasarc.gsfc.nasa.gov/docs/nicer/}, eROSITA
\footnote{extended ROentgen  Survey with an Imaging Telescope Array, http://www.mpe.mpg.de/eROSITA} and LOFT\footnote{Large Observatory For X-ray Timing; 
http://sci.esa.int/loft} and precise timing of binary millisecond pulsars will constrain the super-dense matter equation-of-state (EoS)  \citep{watts2014} and further 
enrich the ensemble of laboratories for strong-field gravity \citep{shao2014}. In $\gamma-$rays, pulsar searches in \emph{Fermi}\footnote{http://fermi.gsfc.nasa.gov} 
and CTA\footnote{Cherenkov Telescope Array, https://www.cta-observatory.org} unidentified sources will allow a better understanding of the NS environment, pulsar 
emission  and binary evolution. Beyond the electromagnetic spectrum, Advanced-LIGO\footnote{https://www.advancedligo.mit.edu}, VIRGO\footnote{http://www.ego-
gw.it/public/about/whatis.aspx}, eLISA\footnote{https://www.elisascience.org} and the Pulsar Timing Array 
(PTA) monitored by the SKA will open a new window to the GW Universe.

In this chapter we elaborate on a selected number of  topics, for which coordination between different observatories will provide the greatest benefits. The text is 
organized as follows: Section\,2 covers the Galactic structure and content, focusing on studies of the Milky Way's kinematics and  multi-wavelength pulsar surveys. 
Section\,3 discusses the added benefits for stellar evolution and NS population studies. In section\,4 we elaborate on the NS EoS and the nature of the Strong 
Interaction and in section\,5 on NS-related transient phenomena.  Finally, section\,6 covers the multi-messenger efforts in the GW detection era and section\,7 
concludes with some final remarks. Given space limitations, this chapter is not meant to be an in-depth review of each topic nor does it exhaust the complete potential 
of the multi-wavelength approach. For details on specific topics the reader is encouraged to skim through the other chapters of this book cited throughout the text.

\section{The Galactic Structure and Content} 
\subsection{Targeted multi-wavelength searches for pulsars}% (Lucas)}

One of the great achievements of \emph{Fermi} was the discovery by its main instrument, the Large Area Telescope (LAT), of a large number of $\gamma$-ray sources 
with no previously known counterparts, the so-called unassociated sources. The 2FGL catalogue \citep{2FGL}, a catalogue of \emph{Fermi} LAT sources based on two 
years of data, contained 1873 sources in total, of which about 30\% have not yet been associated with a known class of gamma-ray source; the recently published 3FGL catalogue, 
which is based on four years of LAT data, contains a total number of sources close to 3000. Many of these could be $\gamma-$ray pulsars, the 
most numerous class of Galactic $\gamma-$ray sources. Indeed, searches for pulsations at the locations of LAT unassociated sources  with pulsar-like $\gamma-$ray 
emission properties have led to the discovery of many new pulsars,  either by directly blind searching the photon data \citep{pletsch2012}, or by conducting deep radio 
observations \citep{ray2012}. 

Radio searches of pulsations in LAT sources have led to the discovery of a very large number of previously unknown MSPs \citep{ray2012} -- currently about 25\% of all 
known in the Galaxy\footnote{See http://astro.phys.wvu.edu/GalacticMSPs/GalacticMSPs.txt for an up-to-date list of known Galactic disk MSPs.}.  A good fraction of 
these new pulsars would probably have \emph{eventually}  been found in standard radio-pulsar surveys.  However, the  LAT accelerated their discovery  by 
showing radio telescopes where to look.  Until the end of its mission, \emph{Fermi} will continue to  discover new sources and the SKA will be extremely useful in the 
quest for the  identification of the unassociated ones. Similarly,  other instruments covering different wavelengths that have started operating (e.g., GAIA or ALMA)  or 
will start in the future (e.g., LSST,  eROSITA, or  the CTA),  will find new sources across the spectrum  that could be searched for pulsars with the SKA.   To quote some 
examples: faint, variable optical stars detected by GAIA could point to   white dwarfs orbiting unknown  radio pulsars  \citep[e.g.][]{Antoniadis2013},   or ``black widow'' 
systems, with millisecond pulsars   ablating their companion stars with their strong particle winds,   generating optical emission modulated at the orbital period  
\citep[e.g.][]{Romani2012a}.   Another example is the possibility to search  for young radio-emitting pulsars  with the SKA, at the  locations of supernova remnants,  pulsar 
wind nebulae (PWNe) or  unassociated sources  discovered by CTA,  in its future surveys of the  very high energy $\gamma$-ray sky \citep{Dubus2013}. 

\subsection{Probing the Dynamics and Structure of the Galaxy}% (Roberto; edited by John)}

Regardless of the discovery method, measured positions, distances and proper motions of radio pulsars, have allowed to broadly outline the structure of the Galactic 
disk and, among others, unveiled the presence of a warp \citep{yusifov2004}. However, to date, out of $\sim$2200 known radio pulsars, only $\sim$150 (i.e. about 7\%) 
have   measured proper motions and  even less have measured parallaxes,  fundamental in determining their actual location in the Galaxy.

The large number of pulsar discoveries expected by the SKA, both in Phase-1 and later in Phase-2 \citep{keane2014,hessels2014}, and the manifold increase in sensitivity, will 
improve the accuracy of previous results and allow to look for new features like humps or depletions in the Galactic disk that might have escaped previous searches. 
Furthermore, binary pulsars with optical counterparts, where both the proper motion and systemic radial velocity can be measured, will provide  their full three-
dimensional motion in the Galaxy \citep{Lazaridis2009} enabling the reconstruction of the Milky-Way's structure and possibly unveiling thin and thick disk pulsar 
populations. Similarly, precise knowledge of pulsar proper motions \citep{smits2011} will allow to discriminate between pulsars born in the warped regions of the Galaxy 
from those born in the Galactic plane. In all these respects, the SKA will complement the work done by GAIA which will sample different stellar populations.

\subsection{The Interstellar and Intergalactic Medium}% (Pablo; edited by John)}

In addition to the above, effects like dispersion measure, Faraday rotation, scattering and scintillation, routinely measured in pulsar observations, provide valuable 
information about the interstellar medium (ISM). Since pulsars are typically very faint, until now, mostly those located in the vicinity of the Sun have been used for 
studies of the ISM \citep{NE2001}. The superb sensitivity of  the SKA in Phase\,II will  extend precision ISM studies to greater distances \citep{han2014} and further enable 
studies of the intergalactic medium (IGM) through the detection of pulsars and fast radio bursts (FRBs) in other galaxies. 

If the distance to the host galaxy is known by e.g. optical and near-infrared studies of standard candles and Cepheid variables, the electron column density and  magnetic field parallel to the line-of-sight to the pulsar can be calculated. These distances to nearby galaxies have been historically difficult to measure with precision \citep{Jacoby1992}, but the new and more powerful optical and infrared telescopes such as E-ELT and JWST will significantly improve distance measurements.

\section{Extreme Astrophysics and Stellar Evolution}  
\subsection{Studies of the pulsar emission mechanism across the electromagnetic spectrum}% (Pablo & Lucas, edited by John)}
While the general concept of pulsar electromagnetic emission  is fairly well established, the complex details of the radiation processes such as the exact emission 
heights and the relevant importance of various emission mechanisms are still unclear. Different approaches and models attempting to
explain the geometry and physical processes responsible for pulsar emission rely on a broad range of assumptions that lead to different predictions
 \citep{LyneGrahamSmith2012}.  The different techniques 
utilized to study the physical processes in emission regions, e.g., profile shape studies and polarimetry, make use of information at 
multiple frequencies across the radio spectrum \citep{LyneGrahamSmith2012}. As the radiation from pulsars is broadband, and produced presumably by several distinct 
radiative 
mechanisms, the coverage of the multiple wavelengths across the electromagnetic spectrum is required to build a complete analysis of the problem. With the SKA and 
the new  ground and space borne observatories covering virtually the entire electromagnetic spectrum with unprecedented sensitivity and time resolution, a much 
better understanding of the emission processes from pulsars is possible. This new era of cutting-edge instrumentation may answer many of the related questions, that 
last now more than 45 years. 

Beyond that, population analyses of $\gamma-$ray pulsars will yield information on the fraction of ``Geminga-like'' pulsars: pulsars that are only visible from high-
energy observations or with extremely low radio luminosities, presumably because the radio emission beams do not cross or only graze our line of sight. The ratio of 
radio-loud to radio-quiet pulsars is a key observable of high-energy emission models \citep{WattersRomani2011} and the SKA's great sensitivity will be particularly 
useful for constraining this ratio, and thus understanding pulsar emission across the spectrum. 

\subsection{Pulsar Wind Nebulae}
There will also be significant synergies between  the SKA 
and upcoming high-energy facilities like the Cerenkov Telescope
Array (CTA) regarding the study of PWNe.
Powered by the rotational energy of their central NS, these
objects are detected across the electromagnetic spectrum, and
currently dominate the Galactic population of TeV $\gamma$-ray
sources \citep[e.g.][]{hartman1999}.  The radio emission from PWNe is believed to be synchrotron
emission from the electrons and positrons created in the pulsar
magnetosphere interacting with the PWN's magnetic field, while its
$\gamma$-ray emission is believed to result from these high-energy
lepton inverse Compton scattering of background photons.  Detecting
both the radio and $\gamma$-ray emission from a PWN allows one to
measure the electron particle spectrum, magnetic field strength, and
energy density of the background photon field -- critical for
understanding the generation and acceleration of leptons in these
objects \citep{gaensler2006,gelfand2009}.  However, such an analysis is currently possible for only 
few sources, since many TeV PWNe remain undetected at radio wavelengths.
This is likely the result of a low magnetic field strength and large
angular size, resulting in a radio surface brightness too low to be
detected with current facilities.  However, the significant
improvement in sensitivity of the SKA, especially on large angular
scales, will allow us to detect radio emission from existing TeV PWNe
as well as any new PWN candidate detected by CTA.  Additionally, the
SKA will also have the sensitivity to discover PWN and pulsars in
currently unidentified TeV sources \citep{gel}.  Together, the CTA and SKA have
the potential for revolutionizing our understanding of these sources.

\subsection{A multi-wavelength view of stellar evolution}% (John)}

The advent of sensitive multi-wavelength surveys described above will uncover a diverse population of binary NSs. This rich NS ensemble will greatly increase the 
chances for revision, and ultimately unification of the stellar formation and evolution paradigm \citep{tauris2014}. For example, the SKA will enable extremely precise 
mass measurements for a large number of pulsars in binaries, allowing a thorough statistical study of the mass-transfer mechanics and the distribution of NS masses at 
birth \citep{kiziltan2013,ozel2012}. At the same time, sensitive optical instruments such as GAIA, LSST and E-ELT will measure the radial velocities, atmospheric 
composition, proper motions and parallaxes for several pulsar companions with optical counterparts, providing further information for the evolutionary history of the 
systems \citep[e.g.][]{antoniadis2012}. Furthermore, as described above, a joint optical/radio effort might increase the chances for finding unique systems. Just to give an 
example, intermediate-mass binary pulsars may be ideal places to look for faint, ultra-cool white dwarfs, which constraint the stellar formation history of the Milky Way 
and might contribute to its ``dark'' baryonic content \citep{kaplan2014}. 
Similarly, X-ray observatories such as LOFT may help to identify more low-mass X-ray binary/MSP transition objects which would help to understand the details of 
pulsar recycling models \citep{archibald2009,Patruno2014,tauris2014}. 
Deep radio observations of pulsars discovered in X-ray or $\gamma$-ray blind searches are also key for understanding the NS luminosity distribution. Unlike high-
energy signals, radio waves are dispersed by free electrons in the propagation path. The dispersion measure of pulsars inferred with radio observations yields their 
(approximate) distance \citep{NE2001}, which is impossible to determine from the high-energy observations alone.

\section{Nuclear Physics and the Strong Interaction}% (John)}

NSs are extremely compact objects; denser than any other object in the current Universe, and anything that has ever been since $\sim 3$\,ms after the Big Bang 
\citep{lattimer2012}. Owing to their extreme properties, they are of fundamental importance for studying the nature of the Strong Interaction which dictates the 
behaviour of matter at densities reaching and exceeding the nuclear saturation density \citep{watts2014}. The EoS describing the bulk properties of matter can 
theoretically be inferred from first-principle QCD calculations. Practically however, the complicated many-body interactions at play render this approach unfeasible. 
Over the past few decades, numerous different approximations have been developed, leading to diverse EoS predictions that span a large space of parameters
\citep{lattimer2012}. 

The EoS of cold nuclear matter, and the way it joins up with the EoS of hot matter, uniquely determine several NS observables such as the NS mass-radius relation, 
moment of inertia, cooling rate, maximum spin and maximum mass above which NSs collapse to black holes. For the first time, these observables will be significantly 
constrained in a range of NS populations with the SKA and next-generation X-ray observatories \citep{watts2014}. The SKA will measure masses for several hundreds   of 
binary pulsars and significantly increase the chances for finding rapidly spinning pulsars \citep{watts2014,hessels2014}. Furthermore it will provide, for the first time, a 
direct measurement of the moment-of-inertia for pulsars in relativistic double-NS systems like J0737$-$3039 \citep{watts2014,keane2014,shao2014}. At the same time 
X-ray missions such as LOFT and Athena will provide simultaneous mass and radius measurements for a handful of NSs residing in X-ray binaries and potentially 
measure the cooling rates of nearby, thermally emitting NSs such as Cas\,A \citep{watts2014}.   

Multi-wavelength targeted survey approaches may also significantly speed-up the search for EoS-constraining pulsars, by telling us where to look: fast-spinning 
pulsars for example are energetic and most likely radiate the bulk of their spin-down energy in the form of $\gamma-$rays \citep{keane2014}. Furthermore, there has 
been increasing evidence that ``black-widow'' and ``redback'' binary pulsars that have optical, X-ray and $\gamma$-ray counterparts, might host massive NSs 
\citep{vankerkwijk2011, romani2012}. Today, precise mass measurements in these systems are challenging, mostly due to sensitivity limits of radio and optical 
telescopes.

\section{Transient Phenomena and the Dynamic Sky}% (Roberto and Pablo}

\subsection{Synergies between the SKA and optical telescopes}
The identification of radio transients, such as Rotating Radio Transients (RRATs) and microquasars, through the interaction with the LSST and LOFT 
\citep{Feroci2012,Lazio2014}, or any other {X}-ray sky monitor to fly in the 2020s, will be one of the main science goals of the SKA \citep{fender2014}. Thanks to their 
location in the southern hemisphere, the synergies between LSST and the SKA will be crucial to elucidate the nature of thousands of transients in the restless radio and 
optical sky. 
 With its continuous monitoring of 20000 square degrees of the sky and its different observing cadence, LSST will discover thousands of transient events on time scales 
 ranging from tens of seconds to hours over 9 decades in flux \citep{lsst}. Furthermore, the LSST will be able to respond to targets-of-opportunity (ToOs) from other 
 facilities with a reaction time of 60s in its Rapid Response Mode. Due to its large field of view of almost 10 square degrees, the LSST will provide colour information in 
 six bands for several fast transients at each time, following the light curve evolution before, during, and after the event and provide quick localisation for follow-ups 
 with other facilities \citep{lsst}. Spectral information in the optical (including photometric redshifts for AGNs) will be crucial to complement the spectral coverage in the radio provided by the SKA and elucidate the nature of the transient, discriminating, e.g. a Galactic microquasar from an AGN. 
Inversely, an SKA trigger of a fast radio transient for LSST follow-up may be crucial for determining its nature. 
RRATs and other sort of bursting NSs would be probably undetectable in optical integration much longer than the length of the radio burst (typically a fraction of a 
second for the RRATs \citep{keane2011}), with the signal from a possible optical burst (assuming that it lasts as long as the radio burst) being washed out. While the 
non-detection of a candidate RRAT in the optical down to mag$\sim$24.5 (the typical sensitivity of $2\times15$\,s-long LSST snapshot integration) would still be 
provide information to help identify its NS, more intriguing synergies would emerge with the E-ELT if it is equipped with suited instruments to fully exploit its potentials 
in time-domain astronomy down to the ms time scales. The detection of simultaneous optical and radio  bursts from RRATs in coincidence with the radio burst, a goal 
that we missed so far, will shed light on the origin of these events and will allow to test the proposed models by, e.g. comparing the optical and radio fluence, the 
profile of the burst light curve, and the characteristics of the radio and optical pulsations (typically detected during a RRAT burst), including possible time lags. 

\subsection{Transients with the SKA and X-ray Telescopes}
The real-time identification of Galactic X-ray transients and follow-up of variable sources discovered by LOFT, eROSITA and, a posteriori, by \emph{Fermi} is another 
field that will greatly benefit from the synergy with the SKA. For example LOFT, with its Wide Field Monitor (WFM), which will cover at least 50\% of the sky 
simultaneously in the 2-50 keV energy band,  will be a perfect discovery machine  of X-ray transients, such as magnetars, which have been well known and extensively 
studied for the past three decades. However, well known does not necessarily mean well understood. After the discovery of the prototype source back in 1979, only 30 
magnetars have been found (including candidates) and we are still in a position where better understanding would benefit from the discovery of more sources. The 
recent discovery of both radio-loud \citep[e.g.][]{shannon2013} and low magnetic field magnetars \citep{rea2010}, has triggered a profound rethinking of the nature of 
these objects. Most magnetars have been identified by their transient X-ray emission. Previous large field-of-view telescopes, such as the All-Sky Monitor aboard RXTE, 
were instrumental in this role, spotting several  candidates. With a factor of 20 larger collecting area, the WFM will discover many more magnetar candidates, triggering 
alerts for other new facilities, including the SKA. Following-up magnetars in radio during, but not only, their bursting phase is key to solve the long-standing dichotomy 
on the magnetar radio-quietness or radio-loudness and peer deep into the very nature of magnetars. 

Other types of erratic variability in radio pulsars can be explored by the LOFT and SKA. A few pulsars tend to emit Giant Radio Pulses (GRPs) and, so far, similar 
phenomena at different energies \citep[Giant Optical Pulses;][]{strader2013} have been observed to occur simultaneously in the Crab pulsar. Do they occur in the X rays as 
well? Only the SKA and LOFT will be able to answer this question. Phenomena discovered only recently in radio pulsars are the mode switches observed synchronously in 
X rays and in radio, like e.g. in PSR B0943+10 \citep{Hermsen2013}. How many other radio pulsars show a similar behaviour? The SKA and LOFT will certainly find more 
of such cases.  On longer time scales, the nature of the many variable X-ray sources that eROSITA will discover in its 4 year all-sky survey (8 scans in total) could be 
clarified by the SKA and optical facilities.  

\subsection{Fast Radio Bursts}
Finally, an emerging field where synergy may be proven useful is the discovery and characterization of FRBs.
FRBs are temporary isolated impulsive bursts of radio emission with very short duration of a few milliseconds. An initial discovery by \cite{Lorimer2007} was followed by 
a small number of detections \citep{Keane2012,Thornton2013,Spitler2014} which dissipated the initial doubts about the astrophysical origin of the bursts. The sources 
of FRBs remain elusive but their large measured dispersion measures and sky locations suggest a possible extragalactic origin. If this is the case, apart from the 
importance of discovering a likely new population of astronomical sources at cosmological distances, they will become important tools for the study of the intergalactic 
medium \citep{Ginzburg1973}. The inherent short duration of FRBs make their detection very difficult in current radio surveys, where a big single-dish telescope with its 
small field of view is scanning the whole sky on many-year timescales. Additionally, the commonly adopted off-line processing of survey data usually leads to detection 
of events well after they have occurred on the sky. What makes these phenomena even more intriguing is that no long-lasting counterpart in radio or other wavelengths 
is found in the direction of the detected bursts. This could mean that additional radiation of the burst at other wavelengths, if any, is also short-lived. The SKA's wide 
field of view will allow to monitor big portions of the sky at once, increasing significantly the detection rate for these FRBs, which are thought to occur several thousand 
times-per-day\,per-sky \citep{Thornton2013,Spitler2014}. In addition, the proposed real-time processing back-end for the survey data in search of short radio bursts 
enable the necessary rapid follow-up of the detected events. A real-time warning from an SKA detection could trigger rapid ToO observations at other wavelengths in 
the burst direction. The SKA in combination with the observatories at all other wavelengths will be key to resolve the mystery of the origin of FRBs and let us study in 
detail the sources and the physical processes producing these unexpected bursts of radiation.

\section{Strong-Field Gravity and the Large-Scale Structure}

\subsection{Studies of the Galactic-Centre black hole across the spectrum}% (Michael)}

Observations of the orbits of the so-called ``S-stars'' have provided detailed information about the object in the centre of our Galaxy. The observations provide the 
most convincing case for it being a super-massive black hole \citep{Genzel2010,MeliaFalcke2001}. First detected in the radio as a point source named Sgr A* (Sagittarius 
A*), the source is now being studied also at near-infrared and X-ray wavelengths. Tracing the orbits of the S-stars, one can derive the distance to the Galactic Centre (8 
kpc) and the mass of the black hole (4 million solar masses). Finding (even normal) pulsars orbiting Sgr A*, the spin and the quadrupole moment can be determined 
with high precision \citep{eatough2014}. These measurements can be compared with constraints to be derived with high-precision optical astrometry of the inner-most 
stars. Measuring for instance the mass of Sgr A* with radio pulsars to a precision of one solar mass \citep{eatough2014}, we can determine the distance to the Galactic 
Centre using the optical observation with a precision to about 1 pc, providing a firm anchor for our understanding of the Galactic dynamics.

\subsection{Multi-Messenger Gravitational Wave Science}% (Michael \& Ben)}
Precision timing with the SKA will start an era of GW astronomy with pulsars. Phase I of the SKA will virtually guarantee the detection of a stochastic GW 
background that has emerged from a population of super-massive binary black holes that was present in processes of early galaxy formation \citep{janssen2014}. Phase 
II of the SKA will allow studying this background in great detail, thereby, for instance, providing insight into the fundamental properties of gravitons such as their spin 
and mass \citep{lee2010}. In general, the SMBH population that is detectable with a PTA experiment is of higher mass than those of sources detectable with the space-
based GW detector eLISA. With PTAs being sensitive to SMBHs with $10^7$ solar masses and orbital periods between 10 and 20 years, SKA observations provide a truly 
complementary window to the SMBH population. Moreover, observations with the SKA can provide measurements of the amplitude and spectral shape of the GW 
background, which encodes information about galaxy merger and SMBH accretion processes. As for instance pointed out by \cite{Sesana2013}, the amplitude of the 
signal tracks the number of occurred mergers integrated over the redshift range, while the spectral shape should contain a break frequency where contribution of 
individual systems becomes important. Indeed, some individual sources may produce a signal that is significantly larger than indicated by the average spectrum, 
allowing the detection of a single source which will eventually evolve into the eLISA frequency band.   

In a comprehensive review Burke-Spolaor \cite{bur13} discusses in more detail
the possibilities, and importance, of electromagnetic identification of the sources of GWs,
which could be from continuous-wave, burst or GW memory sources. PTAs are already being
used to dis-prove the identification of supermassive black hole binaries in the local neighbourhood \cite{jllw04}.
As the sensitivity of PTAs improves dramatically with the SKA, improved limits and ultimately detections will
be possible. Localising these sources by use of an electromagnetic counterpart such as might be possible 
with current surveys in optical and X-ray wavelengths and using future facilities like LSST, IXO/Athena, Astro-H
combined with the GW signature will allow significantly more detailed information about the binary system to be 
determined.

For compact relativistic binaries with sufficiently small orbital periods, it is possible that GWs may be directly detectable with eLISA. The orbital frequency of the Double 
Pulsar, for instance, is $1.16 \times 10^{-4}$ Hz, so that we can expect to observe a strain of about $5 \times 10^{-21}$ at $2.3 \times 10^{-4}$ Hz 
\citep{KramerStairs2008}. Realistically, a detection may be aggravated by the large expected background of double-white-dwarf systems with similar orbital periods 
\citep{Nelemans2001}. However, if the orbital ephemerides are well known from radio timing with the SKA, and because the systems should also produce power at the 
next orbital harmonic, it should be possible to detect the appropriate sources in a coherent search that takes advantage of the known direction to the source (S. 
Sigurdsson \& C. Miller, private communication). If detection is made, it is possible to combine the observations obtained from the radio with those obtained with eLISA. 
This combination should in principle be able to provide the exact distance to the source, the true inclination angle of the system (rather than either sine or cosine of the 
inclination angle) and the masses. Therefore, the system should be vastly over-determined, allowing to provide unprecedented tests of theories of gravity.

Isolated neutron stars may also be the source of GWs if they are deformed in such a way as to
make them axi-symmetric. The GW amplitude  will depend on the size of the asymmetry 
which in turn is strongly dependent on the nature of the equation of state of the neutron star and the strength
of the internal magnetic fields. Significant and important limits on the degree of deformation and the fraction of 
the spin-down energy loss of pulsars that might manifest as GW emission have been obtained
with current generation GW observatories \cite[e.g.][]{aaa14,ligo14} however the SKA will be operating at the same
time as the much more sensitive advanced LIGO and VIRGO detectors. To be able to undertake these 
searches for GWs typically involves long integration times and it is therefore necessary to have
good models of the rotational history of the pulsars. The SKA will help by discovering many more pulsars, thus
improving chances of finding even just one deformed source and also be enabling the monitoring of even 
larger numbers of pulsars. Moreover we will also be sensitive to systems that might exhibit GW 
bursts. This is only the tip of what might be possible though synergies between the ground-based gravitational
wave observatories and the SKA.

\section{Conclusions}
In this chapter we elaborated on various aspects of the multi-wavelength, multi-messenger NS science that will be enabled in the SKA era. 
Current simulations show that even SKA\,I can discover a total of about 10000 normal pulsars and perhaps as many as 1800 millisecond pulsars (MSPs), with SKA\,I-
LOW surveying the sky with the Galactic latitude $|b| \geq 5^{\rm o}$, and SKA\,I-MID surveying the sky with the Galactic latitude $|b| \leq 10^{\rm o}$ 
\citep{keane2014}. SKA\,II will provide a complete census of radio pulsars and with its Aperture Array systems should allow for an optimum combination of sensitivity, 
field-of-view and 
number of beams to be able to obtain exceptional cadence on a very large number of sources. These key features  will allow for significant advances in our 
understanding of NSs, which could be further accelerated by coordinated efforts with other next-generation telescopes. As demonstrated here, synergies across the 
electromagnetic-spectrum and beyond could provide a better
understanding of (I) the Galactic structure and content, (II) extreme astrophysics and stellar evolution, (III) nuclear physics and the strong interaction, (IV) transient 
phenomena and (V) strong-field gravity and the large-scale structure of the Universe.

\bibliographystyle{apj}

\bibliography{multilambda_20140512.bib}

\begin{thebibliography}{}
\expandafter\ifx\csname natexlab\endcsname\relax\def\natexlab#1{#1}\fi

\bibitem[{{Aasi} {et~al.}(2014){Aasi}, {Abadie}, {Abbott}, {Abbott}, {Abbott},
  {Abernathy}, {Accadia}, {Acernese}, {Adams}, {Adams}, \& et~al.}]{aaa14}
{Aasi}, J., {Abadie}, J., {Abbott}, B.~P., {et~al.} 2014, \apj, 785, 119

\bibitem[{{Antoniadis} {et~al.}(2013){Antoniadis}, {Freire}, {Wex}, \& {et
  al.}}]{Antoniadis2013}
{Antoniadis}, J., {Freire}, P.~C.~C., {Wex}, N., \& {et al.} 2013, Science,
  340, 448

\bibitem[{{Antoniadis} {et~al.}(2012){Antoniadis}, {van Kerkwijk}, {Koester},
  \& {et al.}}]{antoniadis2012}
{Antoniadis}, J., {van Kerkwijk}, M.~H., {Koester}, D., \& {et al.} 2012,
  \mnras, 423, 3316

\bibitem[{{Archibald} {et~al.}(2009){Archibald}, {Stairs}, {Ransom}, \& {et
  al.}}]{archibald2009}
{Archibald}, A.~M., {Stairs}, I.~H., {Ransom}, S.~M., \& {et al.} 2009,
  Science, 324, 1411

\bibitem[{{Burke-Spolaor}(2013)}]{bur13}
{Burke-Spolaor}, S. 2013, Classical and Quantum Gravity, 30, 224013

\bibitem[{{Cordes} \& {Lazio}(2002)}]{NE2001}
{Cordes}, J.~M., \& {Lazio}, T.~J.~W. 2002, ArXiv Astrophysics e-prints,
  ArXiv:0207156

\bibitem[{{Dubus} {et~al.}(2013){Dubus}, {Contreras}, {Funk}, {Gallant},
  {Hassan}, {Hinton}, {Inoue}, {Kn{\"o}dlseder}, {Martin}, {Mirabal}, {de
  Naurois}, {Renaud}, \& {CTA Consortium}}]{Dubus2013}
{Dubus}, G., {Contreras}, J.~L., {Funk}, S., {et~al.} 2013, Astroparticle
  Physics, 43, 317

\bibitem[{{Eatough} {et~al.}(2014){Eatough}, {Lazio}, {Casanellas},
  {Chatterjee}, {Cordes}, {Demorest}, {Kramer}, {Lee}, {Liu}, {Ransom}, \&
  {Wex}}]{eatough2014}
{Eatough}, R.~P., {Lazio}, T.~J.~W., {Casanellas}, J., {et~al.} 2014,
  PoS(AASKA)045

\bibitem[{{Fender}(2014)}]{fender2014}
{Fender}, R. 2014, PoS(AASKA)051

\bibitem[{{Feroci} {et~al.}(2012){Feroci}, {Stella}, {van der Klis}, \& {et
  al.}}]{Feroci2012}
{Feroci}, M., {Stella}, L., {van der Klis}, \& {et al.} 2012, Experimental
  Astronomy, 34, 415

\bibitem[{{Gaensler} \& {Slane}(2006)}]{gaensler2006}
{Gaensler}, B.~M., \& {Slane}, P.~O. 2006, \araa, 44, 17

\bibitem[{{Gelfand} {et~al.}(2014){Gelfand}, {Breton}, {Ng}, {Hessels},
  {Stappers}, {Roberts}, \& {Possenti}}]{gel}
{Gelfand}, J.~D., {Breton}, R.~P., {Ng}, C.-Y., {et~al.} 2014, PoS(AASKA)046

\bibitem[{{Gelfand} {et~al.}(2009){Gelfand}, {Slane}, \& {Zhang}}]{gelfand2009}
{Gelfand}, J.~D., {Slane}, P.~O., \& {Zhang}, W. 2009, \apj, 703, 2051

\bibitem[{{Genzel} {et~al.}(2010){Genzel}, {Eisenhauer}, \&
  {Gillessen}}]{Genzel2010}
{Genzel}, R., {Eisenhauer}, F., \& {Gillessen}, S. 2010, Reviews of Modern
  Physics, 82, 3121

\bibitem[{{Ginzburg}(1973)}]{Ginzburg1973}
{Ginzburg}, V.~L. 1973, \nat, 246, 415

\bibitem[{{Han} {et~al.}(2014){Han}, {van Straten}, {Lazio}, {Deller}, {Sobey},
  {Xu}, {Schnitzeler}, {Imai}, {Chatterjee}, {Macquart}, {Kramer}, \&
  {Cordes}}]{han2014}
{Han}, J.~L., {van Straten}, W., {Lazio}, T.~J.~W., {et~al.} 2014,
  PoS(AASKA)043

\bibitem[{{Hartman} {et~al.}(1999){Hartman}, {Bertsch}, {Bloom}, \& {et
  al.}}]{hartman1999}
{Hartman}, R.~C., {Bertsch}, D.~L., {Bloom}, \& {et al.} 1999, \apjs, 123, 79

\bibitem[{{Hermsen} {et~al.}(2013){Hermsen}, {Hessels}, \&
  {Kuiper}}]{Hermsen2013}
{Hermsen}, W., {Hessels}, J.~W.~T., \& {Kuiper}, e. 2013, Science, 339, 436

\bibitem[{{Hessels} {et~al.}(2014){Hessels}, {Possenti}, {Bailes}, {Bassa},
  {Freire}, {Lorimer}, {Lynch}, {Ransom}, \& {Stairs}}]{hessels2014}
{Hessels}, J.~W.~T., {Possenti}, A., {Bailes}, M., {et~al.} 2014, PoS(AASKA)047

\bibitem[{{Ivezic} {et~al.}(2008){Ivezic}, {Tyson}, {Acosta}, {Allsman}, {et
  al.}, \& {for the LSST Collaboration}}]{lsst}
{Ivezic}, Z., {Tyson}, J.~A., {Acosta}, E., {et~al.} 2008, ArXiv e-prints,
  arXiv:arXiv:0805.2366

\bibitem[{{Jacoby} {et~al.}(1992){Jacoby}, {Branch}, {Ciardullo}, {Davies}, \&
  {et al.}}]{Jacoby1992}
{Jacoby}, G.~H., {Branch}, D., {Ciardullo}, R., {Davies}, R.~L., \& {et al.}
  1992, \pasp, 104, 599

\bibitem[{{Janssen} {et~al.}(2014){Janssen}, {Hobbs}, {McLaughlin}, {Bassa},
  {Deller}, {Kramer}, {Lee}, {Mingarelli}, {Rosado}, {Sanidas}, {Sesana},
  {Shao}, {Stairs}, {Stappers}, \& {Verbiest}}]{janssen2014}
{Janssen}, G.~H., {Hobbs}, G., {McLaughlin}, M., {et~al.} 2014, PoS(AASKA)037

\bibitem[{{Jenet} {et~al.}(2004){Jenet}, {Lommen}, {Larson}, \& {Wen}}]{jllw04}
{Jenet}, F.~A., {Lommen}, A., {Larson}, S.~L., \& {Wen}, L. 2004, \apj, 606,
  799

\bibitem[{{Kaplan} {et~al.}(2014){Kaplan}, {Boyles}, {Dunlap}, \& {et
  al.}}]{kaplan2014}
{Kaplan}, D.~L., {Boyles}, J., {Dunlap}, B.~H., \& {et al.} 2014, \apj, 789,
  119

\bibitem[{{Keane} {et~al.}(2011){Keane}, {Kramer}, {Lyne}, {Stappers}, \&
  {McLaughlin}}]{keane2011}
{Keane}, E.~F., {Kramer}, M., {Lyne}, A.~G., {Stappers}, B.~W., \&
  {McLaughlin}, M.~A. 2011, \mnras, 415, 3065

\bibitem[{{Keane} {et~al.}(2012){Keane}, {Stappers}, {Kramer}, \&
  {Lyne}}]{Keane2012}
{Keane}, E.~F., {Stappers}, B.~W., {Kramer}, M., \& {Lyne}, A.~G. 2012, \mnras,
  425, L71

\bibitem[{{Keane} {et~al.}(2014){Keane}, {Bhattacharyya}, {Kramer}, {Stappers},
  {Bates}, {Burgay}, {Chatterjee}, {Champion}, {Eatough}, {Hessels}, {Janssen},
  {Lee}, {van Leeuwen}, {Margueron}, {Oertel}, {Possenti}, {Ransom},
  {Theureau}, \& {Torne}}]{keane2014}
{Keane}, E.~F., {Bhattacharyya}, B., {Kramer}, M., {et~al.} 2014, PoS(AASKA)040

\bibitem[{{Kiziltan} {et~al.}(2013){Kiziltan}, {Kottas}, {De Yoreo}, \&
  {Thorsett}}]{kiziltan2013}
{Kiziltan}, B., {Kottas}, A., {De Yoreo}, M., \& {Thorsett}, S.~E. 2013, \apj,
  778, 66

\bibitem[{{Kramer} \& {Stairs}(2008)}]{KramerStairs2008}
{Kramer}, M., \& {Stairs}, I.~H. 2008, \araa, 46, 541

\bibitem[{{Lattimer}(2012)}]{lattimer2012}
{Lattimer}, J.~M. 2012, Annual Review of Nuclear and Particle Science, 62, 485

\bibitem[{{Lazaridis} {et~al.}(2009){Lazaridis}, {Wex}, {Jessner}, \& {et
  al.}}]{Lazaridis2009}
{Lazaridis}, K., {Wex}, N., {Jessner}, A., \& {et al.} 2009, \mnras, 400, 805

\bibitem[{{Lazio} {et~al.}(2014){Lazio}, {Kimball}, {Barger}, \& {et
  al.}}]{Lazio2014}
{Lazio}, J.~W., {Kimball}, A., {Barger}, A.~J., \& {et al.} 2014, \pasp, 126,
  196

\bibitem[{{Lee} {et~al.}(2010){Lee}, {Jenet}, {Price}, {Wex}, \&
  {Kramer}}]{lee2010}
{Lee}, K., {Jenet}, F.~A., {Price}, R.~H., {Wex}, N., \& {Kramer}, M. 2010,
  \apj, 722, 1589

\bibitem[{{Lorimer} {et~al.}(2007){Lorimer}, {Bailes}, {McLaughlin},
  {Narkevic}, \& {Crawford}}]{Lorimer2007}
{Lorimer}, D.~R., {Bailes}, M., {McLaughlin}, M.~A., {Narkevic}, D.~J., \&
  {Crawford}, F. 2007, Science, 318, 777

\bibitem[{{Lyne} \& {Graham-Smith}(2012)}]{LyneGrahamSmith2012}
{Lyne}, A., \& {Graham-Smith}, F. 2012, {Pulsar Astronomy}

\bibitem[{{Melia} \& {Falcke}(2001)}]{MeliaFalcke2001}
{Melia}, F., \& {Falcke}, H. 2001, \araa, 39, 309

\bibitem[{{Nelemans} {et~al.}(2001){Nelemans}, {Yungelson}, \& {Portegies
  Zwart}}]{Nelemans2001}
{Nelemans}, G., {Yungelson}, L.~R., \& {Portegies Zwart}, S.~F. 2001, \aap,
  375, 890

\bibitem[{{Nolan} {et~al.}(2012){Nolan}, {Abdo}, {Ackermann}, {Ajello},
  {Allafort}, {Antolini}, {Atwood}, {Axelsson}, {Baldini}, {Ballet}, \&
  et~al.}]{2FGL}
{Nolan}, P.~L., {Abdo}, A.~A., {Ackermann}, M., {et~al.} 2012, \apjs, 199, 31

\bibitem[{{{\"O}zel} {et~al.}(2012){{\"O}zel}, {Psaltis}, {Narayan}, \& {Santos
  Villarreal}}]{ozel2012}
{{\"O}zel}, F., {Psaltis}, D., {Narayan}, R., \& {Santos Villarreal}, A. 2012,
  \apj, 757, 55

\bibitem[{{Patruno} {et~al.}(2014){Patruno}, {Archibald}, {Hessels}, \& {et
  al.}}]{Patruno2014}
{Patruno}, A., {Archibald}, A.~M., {Hessels}, J.~W.~T., \& {et al.} 2014,
  \apjl, 781, L3

\bibitem[{{Pletsch} {et~al.}(2012){Pletsch}, {Guillemot}, {Allen}, \& {et
  al.}}]{pletsch2012}
{Pletsch}, H.~J., {Guillemot}, L., {Allen}, B., \& {et al.} 2012, \apj, 744,
  105

\bibitem[{{Ray} {et~al.}(2012){Ray}, {Abdo}, {Parent}, \& {et al.}}]{ray2012}
{Ray}, P.~S., {Abdo}, A.~A., {Parent}, D., \& {et al.} 2012, ArXiv e-prints,
  arXiv:ArXiv:1205.3089

\bibitem[{{Rea} {et~al.}(2010){Rea}, {Esposito}, {Turolla}, \& {et
  al.}}]{rea2010}
{Rea}, N., {Esposito}, P., {Turolla}, R., \& {et al.} 2010, Science, 330, 944

\bibitem[{{Romani}(2012)}]{Romani2012a}
{Romani}, R.~W. 2012, \apjl, 754, L25

\bibitem[{{Romani} {et~al.}(2012){Romani}, {Filippenko}, {Silverman}, \& {et
  al.}}]{romani2012}
{Romani}, R.~W., {Filippenko}, A.~V., {Silverman}, J.~M., \& {et al.} 2012,
  \apjl, 760, L36

\bibitem[{{Sesana}(2013)}]{Sesana2013}
{Sesana}, A. 2013, \mnras, 433, L1

\bibitem[{{Shannon} \& {Johnston}(2013)}]{shannon2013}
{Shannon}, R.~M., \& {Johnston}, S. 2013, \mnras, 435, L29

\bibitem[{{Shao} {et~al.}(2014){Shao}, {Stairs}, {Antoniadis}, {Deller},
  {Freire}, {Hessels}, {Janssen}, {Kramer}, {Kunz}, {L{\"a}mmerzahl},
  {Perlick}, {Possenti}, {Ransom}, {Stappers}, \& {van Straten}}]{shao2014}
{Shao}, L., {Stairs}, I.~H., {Antoniadis}, J., {et~al.} 2014, PoS(AASKA14)042

\bibitem[{{Smits} {et~al.}(2011){Smits}, {Tingay}, {Wex}, {Kramer}, \&
  {Stappers}}]{smits2011}
{Smits}, R., {Tingay}, S.~J., {Wex}, N., {Kramer}, M., \& {Stappers}, B. 2011,
  \aap, 528, A108

\bibitem[{{Spitler} {et~al.}(2014){Spitler}, {Cordes}, {Hessels}, \& {et
  al.}}]{Spitler2014}
{Spitler}, L.~G., {Cordes}, J.~M., {Hessels}, J.~W.~T., \& {et al.} 2014, ArXiv
  e-prints, arXiv:ArXiv:1404.2934

\bibitem[{{Strader} {et~al.}(2013){Strader}, {Johnson}, {Mazin}, \& {et
  al.}}]{strader2013}
{Strader}, M.~J., {Johnson}, M.~D., {Mazin}, B.~A., \& {et al.} 2013, \apjl,
  779, L12

\bibitem[{{Tauris} {et~al.}(2014){Tauris}, {Kaspi}, {Breton}, {Deller},
  {Keane}, {Kramer}, {Lorimer}, {McLaughlin}, {Possenti}, {Ray}, {Stappers}, \&
  {Weltevrede}}]{tauris2014}
{Tauris}, T.~M., {Kaspi}, V.~M., {Breton}, R.~P., {et~al.} 2014, PoS(AASKA)039

\bibitem[{{The LIGO Scientific Collaboration} {et~al.}(2014){The LIGO
  Scientific Collaboration}, {the Virgo Collaboration}, {Aasi}, {Abbott},
  {Abbott}, {Abbott}, {Abernathy}, {Acernese}, {Ackley}, {Adams}, \&
  et~al.}]{ligo14}
{The LIGO Scientific Collaboration}, {the Virgo Collaboration}, {Aasi}, J.,
  {et~al.} 2014, ArXiv e-prints, arXiv:1410.8310

\bibitem[{{Thornton} {et~al.}(2013){Thornton}, {Stappers}, {Bailes}, \& {et
  al.}}]{Thornton2013}
{Thornton}, D., {Stappers}, B., {Bailes}, M., \& {et al.} 2013, Science, 341,
  53

\bibitem[{{van Kerkwijk} {et~al.}(2011){van Kerkwijk}, {Breton}, \&
  {Kulkarni}}]{vankerkwijk2011}
{van Kerkwijk}, M.~H., {Breton}, R.~P., \& {Kulkarni}, S.~R. 2011, \apj, 728,
  95

\bibitem[{{Watters} \& {Romani}(2011)}]{WattersRomani2011}
{Watters}, K.~P., \& {Romani}, R.~W. 2011, \apj, 727, 123

\bibitem[{{Watts} {et~al.}(2014){Watts}, {Xu}, {Espinoza}, {Andersson},
  {Antoniadis}, {Antonopoulou}, {Buchner}, {Dai}, {Demorest}, {Freire},
  {Hessels}, {Margueron}, {Oertel}, {Patruno}, {Possenti}, {Ransom}, {Stairs},
  \& {Stappers}}]{watts2014}
{Watts}, A., {Xu}, R., {Espinoza}, C., {et~al.} 2014, PoS(AASKA14)043

\bibitem[{{Yusifov}(2004)}]{yusifov2004}
{Yusifov}, I. 2004, in The Magnetized Interstellar Medium, ed. B.~{Uyaniker},
  W.~{Reich}, \& R.~{Wielebinski}, 165--169

\end{thebibliography}

\end{document}